\documentclass[aps,prd,reprint,showpacs,preprintnumbers,amsmath,amssymb]{revtex4-1}
\usepackage{graphicx}
\usepackage{dcolumn}
\usepackage{bm}
\usepackage{hyperref}
\hypersetup{
	colorlinks=true,
	linkcolor=blue,
	citecolor=blue,
	urlcolor=blue
	}
\usepackage[mathlines]{lineno}
\usepackage{color}
\usepackage{xcolor}

\newcommand{\etapi}{\eta\to{3\pi}}

\newcommand{\etapipm}{\eta\to{\pi^+\pi^-\pi^0}}

\begin{document}


\title{Patterns of $CP$ violation from mirror symmetry breaking in the $\eta\rightarrow\pi^+\pi^-\pi^0$ Dalitz plot}

\author{Susan Gardner}
\email[]{\text{gardner@pa.uky.edu}}
\affiliation{Department of Physics and Astronomy, University of Kentucky,
Lexington, Kentucky 40506, USA}
\author{Jun Shi} 
\email[]{\text{jun.shi@uky.edu}}
\affiliation{Department of Physics and Astronomy, University of Kentucky,
Lexington, Kentucky 40506, USA}


\begin{abstract} 
A violation of mirror symmetry in the $\eta\to\pi^+\pi^-\pi^0$ Dalitz plot has
long been recognized as a signal of $C$ and $CP$ violation. 
Here we show how the isospin of the underlying $C$- and $CP$-violating 
structures can be reconstructed from their kinematic representation in the 
Dalitz plot. Our analysis of the most recent experimental data reveals, for the first time, 
that the $C$- and $CP$-violating amplitude with 
total isospin $I=2$ is much more severely suppressed than that 
with total isospin $I=0$.
\end{abstract} 

\maketitle


\section{Introduction}
The decay $\eta\rightarrow 3\pi$ first came to prominence after the 
observation of 
$K_L \to \pi^+ \pi^-$ decay and the discovery 
of $CP$ violation in 1964~\cite{Christenson:1964fg}, because 
it could be used to test whether $K_L \to \pi^+ \pi^-$ decay was generated by 
$CP$ violation 
in the weak interactions~\cite{Lee:1965hi,Prentki:1965tt}. Rather, 
$CP$ violation could arise from the interference of the 
$CP$-conserving weak interaction with a new, ``strong'' interaction that breaks $C$ and $CP$; 
this new interaction could be identified through 
the appearance of a charge asymmetry in the momentum distribution of $\pi^+$ and $\pi^-$ in 
$\etapipm$ decay~\cite{Lee:1965hi,Lee:1965zza,Nauenberg:1965}. 
Since $\etapipm$ breaks G parity, 
isospin $I$ and/or charge-conjugation $C$ must be broken in order for the process to occur. 
Thus a charge asymmetry could arise from 
the interference of a $C$-conserving, but isospin-breaking 
amplitude with a 
isospin-conserving, but $C$-violating one~\cite{Lee:1965zza}. 
Numerical estimates were made by assuming that 
the isospin-violating contributions 
were driven by electromagnetism~\cite{Lee:1965zza,Nauenberg:1965,Barrett:1965ia}. 
Since that early work, our understanding of these decays 
within the Standard Model (SM) has changed completely: the weak interaction does indeed
break $CP$ symmetry, through flavor-changing transitions characterized by the
Cabibbo-Kobayashi-Maskawa (CKM) matrix. Moreover, 
isospin breaking in the strong interaction, mediated by the up-down quark mass 
difference~\cite{Gross:1979ur,Langacker:1978cf,Leutwyler:1996qg}, is now known to 
provide the driving effect in mediating $\eta \to 3 \pi$ 
decay~\cite{Gasser:1984pr,Anisovich:1996tx,Bijnens:2002qy,Bijnens:2007pr}, with
isospin-breaking, electromagnetic effects 
playing a much smaller role~\cite{Sutherland:1966zz,Bell:1968mi,Baur:1995gc,Ditsche:2008cq}. 

Modern theoretical studies of $\eta\rightarrow 3\pi$ decay focus on a complete 
description of the final-state interactions within the SM, in order to extract 
the isospin-breaking, light-quark mass ratio 
$Q \equiv \sqrt{(m_s^2 - \hat{m}^2)/(m_d^2 - m_u^2)}$, with $\hat{m}=(m_d + m_u)/2$, 
precisely~\cite{Anisovich:1996tx,Bijnens:2002qy,Bijnens:2007pr,Colangelo:2009db,Kampf:2011wr,Guo:2015zqa,Albaladejo:2017hhj,Colangelo:2016jmc,Colangelo:2018jxw}.  There has been 
no further theoretical study of $CP$ violation 
in $\eta \rightarrow\pi^+\pi^-\pi^0$ decay since 1966. Since the $\eta$ meson carries
neither spin nor flavor, searches for new physics in this system possess special features. 
For example, $\eta\rightarrow\pi^+\pi^-\pi^0$ 
decay must be parity $P$ conserving if the $\pi$ and $\eta$ 
mesons have the same intrinsic parity, so that $C$ violation in this process implies
that $CP$ is violated as well. 
There has been, moreover, much effort invested in the possibility of flavor-diagonal 
$CP$ violation via a non-zero permanent electric dipole moment (EDM), which is $P$ and 
time-reversal $T$ violating, or $P$ and $CP$ violating if $CPT$ symmetry is assumed. 
Studies of flavor-diagonal, $C$ and $CP$ violating processes 
are largely lacking.
We believe that the study of the Dalitz plot distribution in 
$\eta\rightarrow \pi^+ (p_{\pi^+}) \pi^- (p_{\pi^-}) \pi^0 (p_{\pi^0})$ decay
is an ideal arena in which to search for $C$ and $CP$ 
violation beyond the SM. 
Were we to plot the Dalitz distribution in terms of the Mandelstam variables
$t \equiv (p_{\pi^-} + p_{\pi^0})^2$ and $u \equiv (p_{\pi^+} + p_{\pi^0})^2$, 
the charge asymmetry we have noted corresponds to a failure of mirror symmetry, i.e., 
of $t\leftrightarrow u$ exchange, in the Dalitz plot. 
In contrast to that $C$ and $CP$ violating observable, 
a nucleon EDM could be mediated by a minimal $P$- and $T$-violating interaction, 
the mass-dimension-four $\bar\theta$ term of the SM, and not new weak-scale physics. 
The $\bar\theta$ term can also generate $\eta\to \pi\pi$ and $\eta/\eta^\prime\to 4\pi^0$ decay,
breaking $P$ and $CP$ explicitly, so that limits on the decay rate constrains
the square of a $CP$-violating 
parameter~\cite{Pich:1991fq,Gutsche:2016jap,Zhevlakov:2018rwo,Guo:2011ir}. Since the
$\bar\theta$ term is $C$ even, it cannot contribute to the charge asymmetry, 
at least at tree level. Moreover, SM weak interactions do 
not support flavor-diagonal $C$ and $CP$ violation. 
Note that the charge asymmetry is linear in $CP$-violating parameters. 

The appearance of a charge asymmetry and thus of $C$ (and $CP$) violation in 
$\eta\rightarrow \pi^+ (p_{\pi^+}) \pi^- (p_{\pi^-}) \pi^0 (p_{\pi^0})$ decay
can be probed experimentally through the measurement of a 
left-right asymmetry, $A_{LR}$~\cite{Layter:1972aq}: 
\begin{equation}
A_{L R} \equiv \frac{N_+ - N_-}{N_+ + N_-} \equiv \frac{1}{N_{\rm tot}} (N_+ - N_-)\,,
\label{asym_LR}
\end{equation}
where $N_{\pm}$ is the number of events with $u\stackrel{>}{{}_{<}}t$, 
so that the $\pi^+$ has 
more (less) energy than the $\pi^-$ if $ u > (<) t$ in the $\eta$ rest system. 
A number of experiments have been conducted over the years to test for a charge asymmetry 
in $\eta\rightarrow\pi^+\pi^-\pi^0$ decay, with early experiments
finding evidence for a nonzero asymmetry~\cite{Baltay:1966zz,Gormley:1970qz,Gormley:1968zz}, 
but with possible systematic problems becoming apparent only later, as, e.g., in 
Ref.~\cite{Anastasi:2016cdz}. 
Other experiments find no evidence for a charge asymmetry and 
$C$ violation~\cite{Larribe:1966zz,Layter:1972aq,Layter:1973ti,Jane:1974mk,Ambrosino:2008ht,Anastasi:2016cdz}, and we note that new, high-statistics experiments are planned~\cite{Gan:2017kfr,Gatto:2016rae,Beacham:2019nyx}.
It is also possible to form asymmetries that probe 
the isospin of the $C$-violating final state: a sextant asymmetry 
$A_S$, sensitive to the $I=0$ state~\cite{Nauenberg:1965,Lee:1965zza}, and 
a quadrant asymmetry $A_Q$, sensitive to the $I=2$ final state~\cite{Lee:1965zza,Layter:1972aq}. 
These asymmetries are more challenging to measure and are only poorly known~\cite{Layter:1972aq}. 
In this paper we develop a method to discriminate between the possible 
$I=0$ and $I=2$ final states by considering the pattern of mirror-symmetry-breaking events
they engender in the Dalitz plot. 
Mirror-symmetry breaking as a probe of $CP$ violation has also been studied 
in untagged, 
heavy-flavor decays~\cite{Burdman:1991vt,Gardner:2002bb,Gardner:2003su,Petrov:2004gs}, with 
Ref.~\cite{Gardner:2003su} analyzing how different $CP$-violating mechanisms populate the 
Dalitz plot. We also note Refs.~\cite{Bediaga:2009tr,Bediaga:2012tm} for Dalitz studies
of $CP$ violation in heavy-flavor decays. 

\section{Theoretical Framework}
The $\etapi$ decay amplitude in the SM 
can be expressed as~\cite{Gasser:1984pr,Anisovich:1996tx}
\begin{eqnarray}
 A (s, t, u) &=& - \frac{1}{Q^2} \frac{M_K^2}{M_{\pi}^2} \frac{M_K^2 -
   M_{\pi}^2}{3 \sqrt{3} F_{\pi}^2} M(s,t,u), 
\label{defA}
  \end{eqnarray}
where we employ the Mandelstam variables $u$, $t$, and $s= (p_{\pi^+} + p_{\pi^-})^2$
and work to leading order in strong-interaction isospin breaking.
Since $C=-(-1)^I$ in $\etapi$ decay~\cite{Lee:1965zza}, the $C$- and $CP$-even 
transition amplitude with a $\Delta I=1$ isospin-breaking prefactor
must have $I=1$. 
The amplitude $M (s, t, u)$ thus 
corresponds to the total isospin $I=1$ component  of the $\pi^+\pi^-\pi^0$ state and 
can be expressed as~\cite{Anisovich:1996tx,Lanz:2018mhm}
  \begin{eqnarray} 
 M^{C}_{1} \!(s, t, u) &=& M_0 (s) + (s - u) M_1 (t) + (s - t) M_1 (u) \nonumber\\
 &+& M_2 (t) + M_2(u) - \frac{2}{3} M_2 (s) \,,
\label{anileut}
  \end{eqnarray}
  where $M_I$(z) is
  an amplitude with $\pi-\pi$ rescattering in the $z$-channel
with isospin $I$. 
This decomposition can be recovered under isospin symmetry 
in chiral perturbation theory (ChPT) up to next-to-next-to-leading order (NNLO), 
${\cal O}(p^6)$,
because the only absorptive parts that can 
appear are in the $\pi-\pi$ $S-$ and $P$-wave amplitudes~\cite{Bijnens:2007pr}. 
An analogous relationship 
exists in $\eta\rightarrow 3\pi^0$ decay~\cite{Anisovich:1996tx}, though 
there is no Dalitz plot asymmetry and hence no effect linear in $CP$ violation in that case 
because the final-state particles are all identical. 

Since we are considering $C$ and $CP$ violation, additional amplitudes 
can appear---namely, total $I=0$ and $I=2$ amplitudes. 
The complete amplitude is thus 
\begin{eqnarray}
A(s,t,u) 
&=& - \frac{1}{Q^2} \frac{M_K^2}{M_{\pi}^2} \frac{M_K^2 -
   M_{\pi}^2}{3 \sqrt{3} F_{\pi}^2} M^{C}_{1} \!(s,t,u) \nonumber \\ 
&& + \alpha M^{\not C}_{0} \!(s,t,u) 
+ \beta M^{\not C}_{2} \!(s,t,u) \,,
\label{total_amplitude}
\end{eqnarray}
where $\alpha $ and $\beta$ are unknown, 
low-energy 
constants---complex numbers to be determined 
by fits to the experimental event populations in the Dalitz plot. 
If they are determined to be nonzero, they signal the appearance of 
$C$- and $CP$-violatiion. 
To construct 
$A_{L R}$ in Eq.~(\ref{asym_LR}), we compute 
\begin{equation}
N_{\pm} = \frac{1}{256 \pi^3 M_\eta^3} \int_{u \stackrel{>}{{}_<}t} dt du \, | A(s, t, u)|^2  \,,
\end{equation}
using Eq.~(\ref{total_amplitude}) and working to leading order in $CP$ violation. 
Since the phase space is symmetric and the $CP$-violating terms are antisymmetric 
under $u \leftrightarrow t$ exchange, we see that 
the $CP$-violating terms leave the total 
decay rate unchanged in ${\cal O}(\alpha)$, ${\cal O}( \beta)$. 

We now turn to the amplitudes $M ^{\not C}_{0,2}(s,t,u)$. 
Here, too, we introduce functions $M_I (z)$ for amplitudes that contain 
 $\pi-\pi$ scattering in the $z$ channel with isospin $I$. 
After using angular-momentum conservation and the Clebsch-Gordon 
coefficients for the addition of the possible isospin states, as shown in the appendix, 
we have 
\begin{equation}
\!\! M^{\not C}_{0} \!(s,t,u) \!= \!(s - t)\! M'_{1}(u) + (u - s)\! M'_{1}(t)
 - (u-t)\! M'_{1}(s)  
\label{CPV_isospin0}
\end{equation}
and
\begin{eqnarray}   
\!\!\!\!\!\!M^{\not C}_{2} \!(s,t,u) \!&=&\! (s - t) M''_1 (u) + (u - s) M''_1 (t) 
\nonumber\\
   && \!\!\!+ 2(u-t)M''_1(s) + \sqrt{5}[M''_2 (u) - M''_2 (t)] \,,
\label{CPV_isospin2}
 \end{eqnarray}    
where the superscripts distinguish the functions that appear in each
state of {\it total} isospin. 
In what follows we do not compute 
$M_I^\prime (z)$ and $M_I^{\prime\prime} (z)$ explicitly, but, rather,
estimate them. 
With this, we can use the experimental studies we consider in this 
paper to set limits on the possibilities, by constraining
$\alpha$ and $\beta$. 
For context  we note that the particular new-physics operators
that would give $C$- and $CP$-violation are not
well established, though examples have been discussed in the
literature~\cite{Khriplovich:1990ef,Conti:1992xn,Engel:1995vv,RamseyMusolf:1999nk,Kurylov:2000ub}. 
From the viewpoint of SM effective field theory (SMEFT) \cite{Buchmuller:1985jz,Grzadkowski:2010es}, we also know that
there are many more examples, even in leading-mass dimension, than have been
discussed thus far~\cite{Gardner:2020}.  Nevertheless, 
we can draw conclusions  
about $M_I^\prime (z)$ and $M_I^{\prime\prime} (z)$ 
irrespective of the choice of new-physics operator. In particular, since 
the operators that mediate $I=0$ or $I=2$ amplitudes break $C$, they cannot 
mediate a $\eta \to \pi^0$ transition, as we suppose that the
neutral meson states remain of definite $C$-parity in the presence of $C$-violation. 
Thus if we were to evaluate the decay diagrams in NLO ChPT in these exotic cases, they would
have the same decay topology as the diagrams that appear in that order in the SM amplitude for
$\eta\to \pi^+\pi^- \pi^0$ decay. 
Thus there is a one-to-one map of the two-body
rescattering terms in the SM to the $C$- and $CP$-violating amplitudes.
To proceed, we assume that the phases of the functions $M_I(z)$, $M_I'(z)$ and $M_I''(z)$ arise
from the strong-interaction dynamics of final-state, $\pi-\pi$ scattering of isospin $I$ in channel z,
making the phase of each function common to all three isospin amplitudes.
Such treatments are familiar from the search for non-SM $CP$ violation,
such as in the study of $B \to \pi (\rho \to \pi \pi)$
decays~\cite{Lipkin:1991st,Snyder:1993mx,Gardner:2001gc,Bigi:2000yz}.
Moreover, at the low energies we consider here, the scattering of the two-pions in the final state
is predominantly elastic, as mixing with other final-states can only occur through G-parity breaking.
Regardless of the total isospin of the final state pions, the effective 
Hamiltonian that mediates the decay separates into a $C$- and/or $I$-breaking piece and a
$C$-  and $I$-conserving piece. Working to leading order in $C$- and/or $I$-breaking, and assuming
that the final-state interactions are two-body only, Watson's theorem~\cite{Watson:1954uc},
familiar from $K \to \pi \pi$ decays~\cite{Bigi:2000yz}, 
also applies to this case and makes the phase of the function $M_I(z)$ common to the three cases.
However, the functions $M_I(z)$, $M_I'(z)$ and $M_I''(z)$ could differ by polynomial prefactors
that depend on $z$. Nevertheless, we believe these effects are relatively negligible, because the
energy release in $\eta \to \pi^+ \pi^- \pi^0$ decay is small. We illustrate this explicitly
later in this section.

We wish to study the possible patterns of $C$- and $CP$-violation across the Dalitz plot, 
so that we now turn to the explicit evaluation of 
Eq.~(\ref{total_amplitude}) and its associated Dalitz distribution. 
Much effort has been devoted to the evaluation of the SM contribution, 
with work in ChPT~\cite{Osborn:1970nn,Gasser:1984pr,Bijnens:2007pr}, 
as well as in frameworks tailored to address various final-state-interaction 
effects~\cite{Neveu:1970tn,Roiesnel:1980gd,Kambor:1995yc,Anisovich:1996tx,Borasoy:2005du,Colangelo:2009db,Schneider:2010hs, 
Kampf:2011wr,Guo:2015zqa,Albaladejo:2017hhj,Colangelo:2016jmc,Colangelo:2018jxw}.  
In what follows we employ a next-to-leading-order (NLO) ChPT 
analysis~\cite{Gasser:1984pr,Bijnens:2007pr}
because it is the simplest choice in which 
the $C$- and $CP$-violating coefficients $\alpha$ and $\beta$ can have both real and imaginary parts. 
A comparison of the NLO and NNLO analyses of 
Bijnens \textit{et al.}~\cite{Bijnens:2007pr}, noting Table I of Ref.~\cite{Anastasi:2016cdz},  
shows that this is an acceptable choice. 
We thus think it is rich enough to give a basic view as to how our idea works. 
To compute the $C$-violating amplitudes, we decompose the $I=1$ amplitude into 
the isospin basis $M_I(z)$. 
As well known~\cite{Bijnens:2002vr,Bijnens:2007pr,Colangelo:2016jmc,Albaladejo:2017hhj}, 
the isospin decompositions involving 
the $\pi-\pi$ rescattering functions $J^r_{P Q} (s)$ are unique, whereas 
the polynomial parts of the amplitude are not, 
due to the relation 
$s + t + u = 3 s_0$, where $s_0=(M_\eta^2 + 2M_{\pi^+}^2 + M_{\pi^0}^2)/3$. 
Thus there are $M_I(z)$ redefinitions that leave the $I=1$ 
amplitude invariant, as discussed in Ref.~\cite{Bijnens:2002vr}. 
However, since we assume that strong rescattering effects dominate $M_I(x)$,  
we can demand that the 
$I=0,2$ amplitudes remain invariant also. 
As a result, 
only the redefinition $M_0 (s) - \frac{4}{3} \delta_1$ and $M_2 (z) + \delta_1$, 
with $\delta_1$ an arbitrary constant, survives.
In what follows we adopt the NLO analyses of Refs.~\cite{Gasser:1984pr,Bijnens:2007pr}, 
and our isospin decomposition of Ref.~\cite{Gasser:1984pr} is 
consistent with that in Bijnens and Ghorbani~\cite{Bijnens:2007pr} --- its
detailed form 
can be found in the information in 
the appendix. 
Small differences
in the numerical predictions exist, however, due to small differences in 
the inputs used~\cite{Gasser:1984pr,Bijnens:2007pr}, 
and we study their impact explicitly.
Returning to the would-be NLO ChPT computation of the total $I=0,2$ amplitudes,
we note that $C$- and $CP$-violating four-quark operators are generated by operators in
mass dimension 8 in SMEFT~\cite{Gardner:2020}, 
so that these amplitudes start beyond ${\cal O}(p^4)$, though
this is not at odds with pulling out a strong rescattering function. 
The $p^2$-dependence of a $C$- and $CP$-violating operator 
from physics beyond the SM would in part  be realized as dimensionless ratios
involving the new physics scale and would appear in the prefactors $\alpha$
or $\beta$ as appropriate.

\begin{figure}[t]
\hspace*{-0.5cm}
\includegraphics[width=0.52\textwidth]{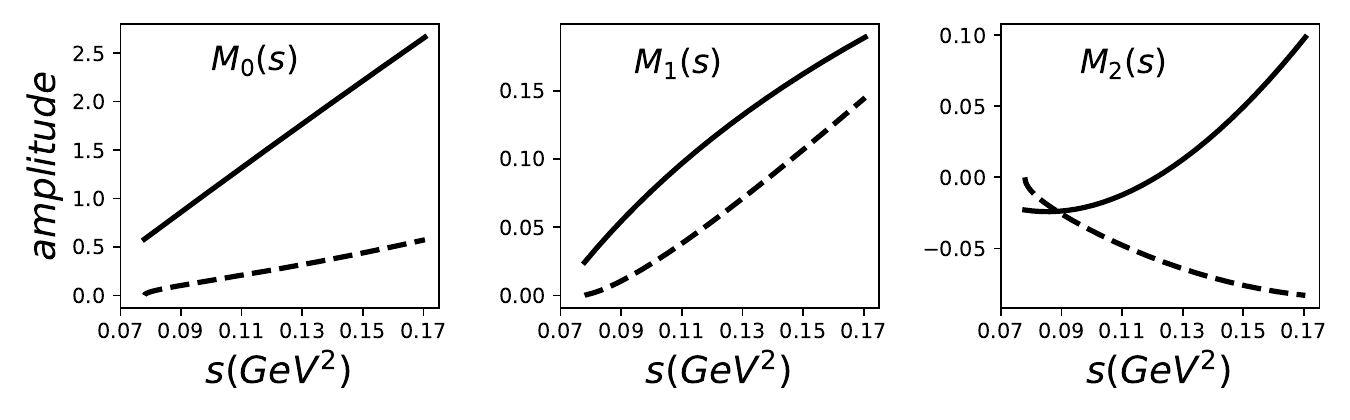}
\caption{Amplitudes of $M_I(s)$ from Eqs.~(\ref{M_0(s)}), (\ref{M_1(s)}) and (\ref{M_2(s)}). 
  The solid lines represent the real part of $M_I(s)$ and the dashed lines
  denote the imaginary part. Their $s$-dependence is driven by that of the 
  $\pi$-$\pi$ phase shift~\cite{Gasser:1984pr}. 
Note that the $M_0(s)$ and $M_2(s)$ amplitudes are dimensionless, whereas 
$M_1(s)$ has units of  ${\rm GeV}^{-2}$.
Since the form $(u-t)M_1(s)$ appears in the final $C$-
and $CP$-violating amplitudes, we note that $M_2(s)$ is typically
a factor of a few larger across the Dalitz plot.  }
\label{fig:M_Is}
\end{figure}

Irrespective of the particular new-physics operator, we note, by analyticity, that the 
$M_I (x)$ for the total $I=0,2$ amplitudes could 
differ from the SM form, which is driven by the strong $\pi-\pi$ phase shifts, 
by a multiplicative polynomial factor, nominally of form $1 + C_1^{I} x/M_{\pi}^2 
+ C_2^{I} x^2/M_{\pi}^4 + \dots$, where $C_1^I$ and $C_2^I$ are constants.
(We note polynomials of similar origin appear in the time-like 
pion form factor~\cite{Gardner:1997ie}.)
We emphasize that in assuming that strong-interaction phases dominate we suppose
these corrections to be unimportant.
We believe this to be an excellent approximation, 
which we illustrate through a plot of the functions $M_I(s)$, as shown 
in Fig.~\ref{fig:M_Is}. The physics of $\pi-\pi$ scattering make the 
functions $M_I(s)$ vary substantially with $s$, whereas $s$ itself only 
changes by about a factor of 2 in $\eta\to 3\pi$ decay. As a result we expect that 
the ignored 
polynomial factors are 
numerically unimportant, so that their neglect 
does not impact the conclusions of this paper. 

\section{Results}
The Dalitz distribution in $\eta \rightarrow \pi^+ \pi^- \pi^0$ 
is usually described in terms of variables 
$X$ and $Y$~\cite{Weinberg:1960zza}: 
\begin{eqnarray}
\!\!\!  X &\equiv& \sqrt{3} \frac{T_{\pi^+} - T_{\pi^-}}{Q_{\eta}} =\frac{\sqrt{3}}{2 M_{\eta} Q_{\eta}} (u - t),\\
\!\!\!  Y &\equiv& \frac{3 T_{\pi^0}}{Q_{\eta}} - 1 = \frac{3}{2 M_{\eta} Q_{\eta}} [(M_{\eta} - M_{\pi^0})^2 - s] - 1\,,
\end{eqnarray}
where 
$ Q_{\eta} = T_{\pi^+} + T_{\pi^-} + T_{\pi^0} = M_{\eta} - 2 M_{\pi^+} -
   M_{\pi^0} $, and $T_{\pi^i}$ is the $\pi^i$ kinetic energy in the $\eta$ rest frame. 
The decay probability 
can be parametrized in a polynomial
expansion around the center point $(X,Y) = (0 ,0)$~\cite{Anastasi:2016cdz}: 
\begin{eqnarray}
\!\!\!\!\! |A (s, t, u)|^2 \!&=&\! N (1 + a Y + b Y^2 + c X + d X^2 + e X Y  \nonumber\\
\!\!\!\!& +& f Y^3 + g X^2 Y + h X Y^2 + l X^3 + \ldots) \,.\label{amplitude_expansion}
 \end{eqnarray}  
Since the $C$ transformation on the decay amplitude
is equivalent to $t \leftrightarrow u$ exchange~\cite{Gardner:2003su}, we see that 
 the appearance of terms that are 
 odd in $X$ 
would indicate both $C$ and $CP$ violation. 
The KLOE-2 collaboration~\cite{Anastasi:2016cdz} has provided a more precise estimate of the 
$C$-even parameters in Eq.~(\ref{amplitude_expansion}) and bounded the $C$-odd ones. 
Returning to Eq.~(\ref{total_amplitude}), we see that the $C$- and $CP$-violating 
contributions to the decay probability 
are
 \begin{eqnarray}
\!\!\!\!\!&& \!\!\!\!\!\!\!\!\frac{1}{\xi}|A(s,t,u)|^2_{\not C}
\!=\! M^C_{1}[\alpha M_{0}^{\not C} + \beta M_{2}^{\not C}]^\ast 
+ H.c. \nonumber\\
\!\!\!\!\!\!\!\!\!\!\!\!\! &=& 
\!2\text{Re}(\alpha)[\text{Re}(M_{1}^C)\text{Re}(M_{0}^{\not C}) 
+\text{Im}(M^C_{1})\text{Im}(M_{0}^{\not C}) ]\nonumber\\
\!\!\!\!\!\! &-&  2\text{Im}(\alpha)[\text{Re}(M^C_{1})\text{Im}(M_{0}^{\not C}) 
-\text{Im}(M_{1}^C)\text{Re}(M_{0}^{\not C}) ]\nonumber\\
\!\!\!\!\!\!&+& 2\text{Re}(\beta)[\text{Re}(M_{1}^C)\text{Re}(M_{2}^{\not C}) 
+ \text{Im}(M_{1}^C)\text{Im}(M_{2}^{\not C}) ]\nonumber\\
\!\!\!\!\!\!&-&  2\text{Im}(\beta)[\text{Re}(M_{1}^C)\text{Im}(M_{2}^{\not C}) 
- \text{Im}(M_{1}^C)\text{Re}(M_{2}^{\not C})]
\,,
 \label{CPV_interference}
 \end{eqnarray}
where $\xi \equiv - (M_K^2/M_{\pi}^2) (M_K^2 - M_{\pi}^2)/(3 \sqrt{3} F_{\pi}^2 Q^2)$, 
and the existing experimental assessments of $|A(s,t,u)|^2_{\not C}$ 
correspond to the set of odd $X$ polynomials in $|A(s, t, u)|^2$. 
The parameter $N$ drops out in the evaluation of the asymmetries, 
and the parameters $c, e, h,$ and $l$ 
are taken from the first line of Table 4.6 in the Ph.D. thesis of  
Caldeira Balkest$\stackrel{\circ}{\text{a}}$hl~\cite{CaldeiraBalkestahl:2016trc}, 
\begin{eqnarray}\label{cehl}
\!\!\!c \!&=& \!(-4.34\pm 3.39)\times 10^{-3}, ~~e \!= \!(2.52\pm 3.20)\times 10^{-3},\nonumber\\
\!\!\!h \!&=& \!(1.07 \pm 0.90)\times 10^{-2}, ~~l \!= \!(1.08\pm 6.54)\times 10^{-3}\,,
\label{liphdcehl}
\end{eqnarray}
which fleshes out Ref.~\cite{Anastasi:2016cdz}---the results emerge from a 
global fit to the Dalitz distribution. 
There is a typographical error in the sign of $c$ in Ref.~\cite{Anastasi:2016cdz}. 
We now turn to the extraction of 
$\mathrm{Re}(\alpha)$, $\mathrm{Im}(\alpha)$, $\mathrm{Re}(\beta)$, and $\mathrm{Im}(\beta)$ 
using the experimental data and 
Eqs.~(\ref{anileut},\ref{CPV_isospin0},\ref{CPV_isospin2}) using the $M_I(z)$ amplitudes from 
${\cal O}(p^4)$ ChPT~\cite{Gasser:1984pr,Bijnens:2007pr}. We evaluate the 
denominators of the possible charge asymmetries by computing $\xi^2 |M_1^C (z,t,u)|^2$ only. 

Herewith we collect the parameters needed for our analysis. 
We compute the phase space with physical masses, so that $s+t+u=3 s_0$, 
but the decay amplitudes~\cite{Gasser:1984pr,Bijnens:2007pr} on which we rely, namely, 
$M(s,t,u)$ in Eq.~(\ref{defA}), should be in the isospin limit, implying 
some adjustment of the input parameters may be needed. We adopt 
the hadron masses and $\sqrt{2}F_\pi=(130.2\pm 1.7)\times 10^{-3}~ \text{GeV}$ 
from Ref.~\cite{Tanabashi:2018} for both amplitudes. 
For the Gasser and Leutwyler (GL) amplitude~\cite{Gasser:1984pr}  we use 
$M_\pi \equiv \sqrt{(2M_{\pi^\pm}^2 + M_{\pi^0}^2)/3}$, 
$M_K \equiv \sqrt{(M_{K^{+}}^2 + M_{K^0}^2)/2}$, where we discuss 
our treatment of the two-particle thresholds in the supplement, with 
$F_0 = F_\pi$, $F_K/F_\pi = 1.1928\pm0.0026$~\cite{Tanabashi:2018}, 
and $L_3= (-3.82 \pm 0.30) \times 10^{-3}$ from the NLO fit 
with the scale $\mu=0.77~ \text{GeV}$~\cite{Bijnens:2014lea}. 
We use these parameters in the prefactor 
in Eq.~(\ref{defA}) also, as well as $Q=22.0$~\cite{Colangelo:2016jmc}, 
to find $\xi=-0.137$. 
For the Bijnens and Ghorbani (BG) amplitude through 
${\cal O}(p^4)$~\cite{Bijnens:2007pr}, we use 
$M(s,t,u) = M^{(2)}(s) + M^{(4)}(s,t,u)$ and multiply the prefactor in Eq.~(\ref{defA}) 
by $-(3F^2_\pi)/(M_\eta^2 - M_\pi^2)$
to yield that in Ref.~\cite{Bijnens:2007pr}. 
In the ${\cal O}(p^2)$ term, which contributes to $M_0(s)$, 
$M^{(2)}(s)= (4M_\pi^2 - s)/F_\pi^2$, and 
we use $M_\pi$ and $F_\pi$ as defined for the GL amplitude~\cite{Gasser:1984pr}. 
In the ${\cal O}(p^4)$ term, we use $M_{\pi^0}$ and $M_{K^0}$ as indicated, 
as well as 
$\Delta=M_\eta^2 - M_{\pi^0}^2$ and $L_3$, $L_5$, $L_7$, $L_8$ from fit 10 of 
Ref.~\cite{Amoros:2001cp}. 

We solve for $\alpha$ and $\beta$ in two different ways for each of the decay 
amplitudes~\cite{Gasser:1984pr,Bijnens:2007pr}. We begin with the GL 
amplitude~\cite{Gasser:1984pr}, 
first making a Taylor expansion of Eq.~(\ref{CPV_interference})
to cubic power in $X$ and $Y$ about $(X,~Y)=(0,~0)$. 
We then equate coefficients associated with the 
$X$, $XY$, $XY^2$, and $X^3$ terms to $c$, $e$, $h$ and $l$, 
respectively, and then solve the four 
equations to obtain 
$\text{Re}(\alpha)$,  $\text{Im}(\alpha)$, $\text{Re}(\beta)$, and $\text{Im}(\beta)$. 
The resulting values of $\alpha$ and $\beta$ are 
\begin{eqnarray}
\text{Re}(\alpha) &=&  16\pm 24 \,,\nonumber\\
\text{Re}(\beta) &=& (-1.5\pm 2.7)\times 10^{-3}\,, \nonumber\\
\text{Im}(\alpha) &=& -20\pm 29\,, \nonumber\\
\text{Im}(\beta) &=& (-1.3\pm 4.7)\times 10^{-3}\,.\label{solution_Taylor}
\end{eqnarray}

\begin{figure}[t]
\vspace*{-0.2cm}
\hspace*{-0.5cm}
\includegraphics[width=0.56\textwidth]{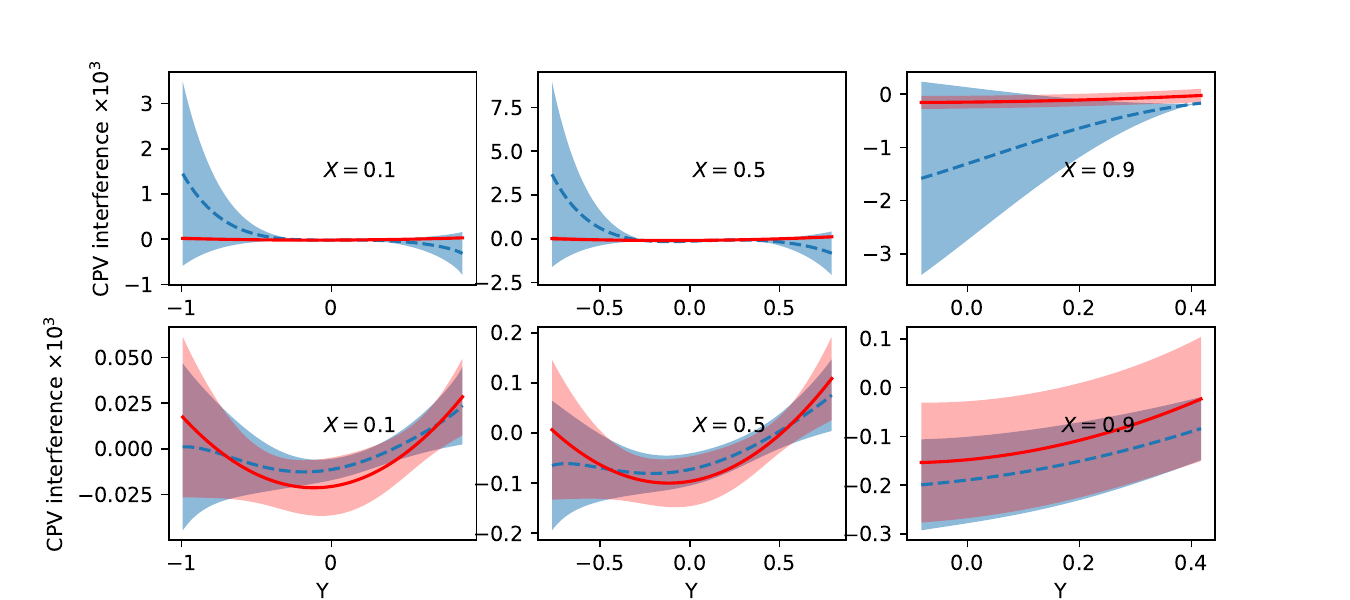}
\vspace*{-0.2cm}
\caption{Results for the $C$- and $CP$-violating (CPV) interference term, 
$|A(s,t,u)|^2_{\not C}$ in Eq.~(\ref{CPV_interference}), 
using the GL amplitude~\cite{Gasser:1984pr} and 
two methods for the determination of $\alpha$ and $\beta$: 
(i) a Taylor expansion (top row) and (ii) a global fit (bottom row) as described in text
using the GL decay amplitude~\cite{Gasser:1984pr}. 
The blue dashed lines with a one-$\sigma$ error band (dark) are our results, and 
the red solid lines with a one-$\sigma$ error band (light) are the KLOE-2 results, as per 
Eq.~(\ref{cehl})~\cite{CaldeiraBalkestahl:2016trc}.}
\label{fig:Taylor_fit}
\end{figure}

In the first row of Fig.~\ref{fig:Taylor_fit} we compare the 
resulting assessment of Eq.~(\ref{CPV_interference}) with the KLOE-2 results. 
Large discrepancies exist, particularly at large values of $X$ and/or $Y$, 
where the empirical Dalitz plot~\cite{Anastasi:2016cdz} shows considerable stength. 
Thus we turn to a second procedure, in which we make a global fit of 
$\alpha$ and $\beta$ in Eq.~(\ref{CPV_interference}) to the KLOE-2 results. 
That is, we assess the Dalitz distribution $N(X,Y)$ and its error by 
using the Dalitz plot parameters in Eq.~(\ref{cehl}), 
discretized onto a $(X,Y)$ mesh with 
682 points. To determine  $N(X,Y)$ and its error we use the odd-$X$ terms in 
Eq.~(\ref{amplitude_expansion}) with the normalization factor $N=0.0474$
as per the GL amplitude~\cite{Gasser:1984pr} and compute the 
covariance matrix using Eq.~(\ref{liphdcehl}) and 
the correlation matrix given in Table 4.3 of Ref.~\cite{CaldeiraBalkestahl:2016trc}. 
We then fit $|A(s,t,u)|^2_{\not C}$ using the GL amplitude~\cite{Gasser:1984pr} 
to $N(X,Y)$ using a $\chi^2$ optimization to find 
\begin{eqnarray}
\text{Re}(\alpha) &=&  -0.65\pm 0.80\,,\nonumber\\
\text{Im}(\alpha) &=& 0.44\pm 0.74\,, \nonumber\\
\text{Re}(\beta) &=& (-6.3\pm 14.7)\times 10^{-4}\,,\nonumber\\
\text{Im}(\beta) &=& (2.2\pm 2.0)\times 10^{-3}\,,\label{solution_fit}
\end{eqnarray}
and we show the results of this method in the second row 
of Fig.~\ref{fig:Taylor_fit}. 
Enlarging the $(X,Y)$ mesh
to 1218 points incurs changes within $\pm1$ of the last significant figure. 
The comparison with experiment 
shows that the fitting procedure is the right choice. 
We draw the same conclusion
from the use of the BG amplitude~\cite{Bijnens:2007pr}, noting that the global
fit in that case (with $N=0.0508$) gives 
\begin{eqnarray}
\text{Re}(\alpha) &=&  -0.79\pm 0.91\,,\nonumber\\
\text{Im}(\alpha) &=& 0.61\pm 0.93\,, \nonumber\\
\text{Re}(\beta) &=& (-1.4\pm 2.3)\times 10^{-3}\,,\nonumber\\
\text{Im}(\beta) &=& (2.3\pm 1.4)\times 10^{-3}\,,\label{solution_fitBG}
\end{eqnarray}
so that the results are compatible within errors. 
Using these solutions, we obtain $A_{LR}=(-7.18\pm 4.51)\times 10^{-4}$ 
using Ref.~\cite{Gasser:1984pr} and $A_{LR}=(-7.20\pm 4.52)\times 10^{-4}$ using
Ref.~\cite{Bijnens:2007pr}. These compare favorably with 
$A_{LR}=(-7.29\pm 4.81)\times 10^{-4}$ that we determine 
using the complete set of Dalitz plot parameters and the covariance matrix we
construct given the information in Ref.~\cite{CaldeiraBalkestahl:2016trc}. 
We note that our $A_{LR}$ as evaluated 
from the Dalitz plot parameters, 
which are fitted from the binned data, is a little different from the reported 
value using the unbinned data, i.e., 
($-5.0\pm 4.5 \stackrel{+5.0}{{}_{-11}})\times 10^{-4}$, reported by 
KLOE-2~\cite{Anastasi:2016cdz}. The discrepancy is not 
significant, 
and we suppose its origin could arise from the slight mismatch between the 
theoretically accessible phase space and the experimentally probed one, or 
other experimental issues. Although NLO ChPT 
does not describe the $CP$-conserving Dalitz distribution well~\cite{Bijnens:2007pr}, 
we find it can confront the existing $CP$-violating observables successfully. 

\begin{figure}[ht]
\centering
\vspace*{-0.2cm}
\includegraphics[width=0.5\textwidth]{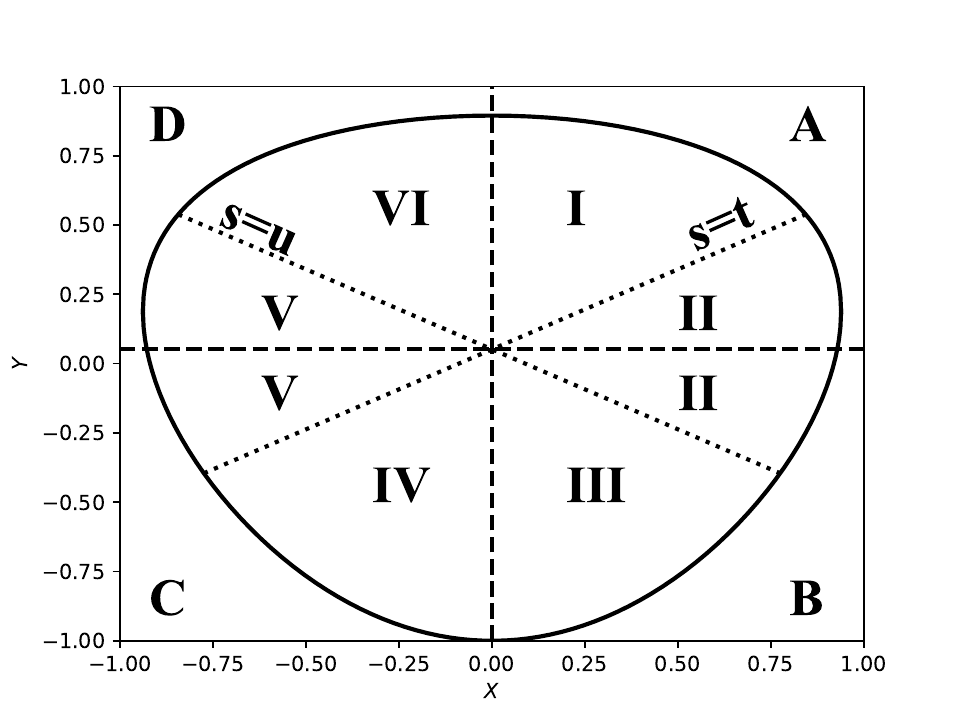}
\vspace*{-0.2cm}
\caption{The Dalitz plot geometry in $\etapipm$ decay, 
 broken into regions for probes of its symmetries. 
The solid line is the boundary of the physically accessible region. 
The asymmetry $A_{LR}$, Eq.~(\ref{asym_LR}), compares the population $N_+$ ($X>0$) 
with $N_-$ ($X<0$). The quadrant asymmetry $A_Q$ probes $I=2$ contributions, 
$N_{\rm tot} A_Q \equiv N({\rm A}) + N({\rm C}) - N({\rm B}) - N({\rm D})$~\cite{Lee:1965zza}, 
and the sextant asymmetry $A_S$ probes $I=0$ contributions, 
$N_{\rm tot} A_S \equiv N({\rm I}) + N({\rm III}) + N({\rm V}) - N({\rm II}) 
- N({\rm IV}) - N({\rm VI})$~\cite{Nauenberg:1965,Lee:1965zza}. 
All asymmetries probe $C$ and $CP$ violation. 
}\label{fig:Dalitz}
\end{figure}

We have shown that 
the empirical Dalitz plot distribution can be used to determine $\alpha$ and $\beta$. 
These, in turn, limit the strength of 
$C$-odd and $CP$-odd operators that can arise from physics beyond the 
SM~\cite{Khriplovich:1990ef,Conti:1992xn,Engel:1995vv,RamseyMusolf:1999nk,Kurylov:2000ub,Gardner:2020}. 
That 
$\beta$ is so much smaller than $\alpha$ can be, in part, understood from the 
differing behavior of
the $M_I(z)$, as illustrated in Fig.~\ref{fig:M_Is}, which
follows
because the
$L=0$, $I=2$ $\pi-\pi$ phase shift is larger than the $L=1$, $I=1$ one
for the kinematics of
interest \cite{Hyams:1973zf,Gasser:1984pr,Colangelo:2001df,Ananthanarayan:2000ht},
making it easier to veto the $I=2$ operators.
Crudely, the ratio of $\beta$ to $\alpha$ we have found is 
that of the SM electromagnetic interactions that would permit a $I=2$ amplitude
to appear in addition to a $I=0$ one. 
The utility of our Dalitz analysis is underscored by our results for the 
quadrant asymmetry $A_Q$ and sextant asymmetry $A_S$ defined in Fig.~\ref{fig:Dalitz}. 
Using Ref.~\cite{Gasser:1984pr} and Eq.~(\ref{solution_fit}), e.g., 
we find 
 $A_Q=(2.85\pm 3.72)\times 10^{-4}$, and $A_S = (3.87\pm 4.04)\times 10^{-4}$; the 
asymmetries by themselves hide the nature of the underlying strong amplitudes. 
For reference we note the KLOE-2 results using unbinned data~\cite{Anastasi:2016cdz}: 
 $A_Q=(1.8\pm 4.5 \stackrel{+4.8}{{}_{-2.3}})\times 10^{-4}$ 
and $A_S = (-0.4 \pm 4.5 \stackrel{+3.1}{{}_{-3.5}})\times 10^{-4}$, with 
which our results are compatible within errors. 

\section{Summary}
We propose an innovative way of probing 
$C$- and $CP$-violation 
in the $\eta\rightarrow\pi^+\pi^-\pi^0$ Dalitz plot.
Working to leading order 
in charge conjugation $C$ and isospin $I$ breaking, we have shown that the 
strong amplitudes associated with the appearance of 
$C$- and $CP$-violation can be 
estimated 
from the SM amplitude for $\etapipm$ 
if the decomposition of Eq.~(\ref{anileut}) holds~\cite{Anisovich:1996tx}. 
We have illustrated this in NLO ChPT, though the use of more sophisticated 
theoretical analyses would also be possible. New-physics contributions
that differ in their isospin can thus be probed through the kinematic pattern they imprint
in the Dalitz plot. 
Our method opens a new window on the study of $C$- and $CP$-violation in 
$\etapipm$ decay, and it 
holds promise for the high-statistics
experiments of the future~\cite{Gan:2017kfr,Gatto:2016rae,Beacham:2019nyx}.

\section{Acknowledgments}
We thank Jian Liang for his assistance in the numerical analysis, 
Johan Bijnens for helpful correspondence regarding Ref.~\cite{Bijnens:2007pr}, 
Andrzej Kupsc for a discussion of the KLOE-2 experimental 
results~\cite{Anastasi:2016cdz,CaldeiraBalkestahl:2016trc}, and Bastian Kubis
for comments that we found helpful to clarifying our presentation. 
We acknowledge the
DOE Office of Nuclear Physics under contract DE-FG02-96ER40989 for partial support.

\appendix*
\section{Calculation Details}  

We begin by showing how Eqs.~(\ref{CPV_isospin0}) and (\ref{CPV_isospin2}) emerge 
from elementary considerations. Working in the isospin limit, 
a $|\pi^+ \pi^- \pi^0\rangle$ state with $J=0$ must obey Bose symmetry, so that 
it is proportional to 
\begin{equation}
|\pi^+ \pi^- \rangle_{\rm s}  |\pi^0 \rangle + 
|\pi^+ \pi^0 \rangle_{\rm s} |\pi^- \rangle + |\pi^- \pi^0 \rangle_{\rm s} | \pi^+ \rangle \,,
\end{equation}
where ``s'' denotes a symmetrized combination of distinct pion states.
In what follows, as in the $CP$-conserving case, Eq.~(\ref{anileut}) ~\cite{Anisovich:1996tx}, we consider 
$S$- and $P$-wave $\pi-\pi$ amplitudes only. The symmetrized two-pion states can be written as a 
$S$-wave $I=0$ or $I=2$ state or as a $P$-wave $I=1$ state. We choose 
$|\pi^i \rangle \equiv | I=1, I_3=i\rangle$. 
For S-waves, we write 
\begin{equation}
|\pi^i \pi^j \rangle_{\rm s} 
\equiv \frac{1}{\sqrt{2}} \{ | \pi^i \pi^j \rangle + | \pi^j \pi^i \rangle  \} \,, 
\end{equation}
whereas for P-waves we note 
\begin{equation}
\underbrace{| \pi^i \pi^j \rangle_{I=1, L=1} | \pi^k \rangle}_{L=1} 
\end{equation}
with $i+ j + k =0$ contributes to the $J=0$ state. Defining 
\begin{equation}
|\pi^i \pi^j \rangle_{\rm a} 
\equiv \frac{1}{\sqrt{2}} \{ |\pi^i \pi^j \rangle - |\pi^j \pi^i \rangle  \} \,, 
\end{equation} 
we see, e.g., 
\begin{equation}
|(\pi^+ \pi^-)_{I=1}\rangle_{\rm s} |\pi^0\rangle 
= |\pi^+ \pi^-\rangle_a  
(p_{\pi^+} - p_{\pi^-}) \cdot p_{\pi^0} |\pi^0\rangle \,.
\end{equation}
We can also label particular 
$\eta \to \pi^+ \pi^-  \pi^0 $ 
decay amplitudes by the isospin of the two-pion state, 
as used in Eq.~(\ref{anileut}). Enumerating the possibilities, we find 
\begin{equation}
|(\pi^+\pi^-)_{I=0} \rangle | \pi^0 \rangle \rightarrow M_0(s) \,,
\end{equation}
which contributes to the total $I=1$ amplitude, $M_1^C$, only, 
as well as 
\small
\begin{eqnarray}
&& |(\pi^+\pi^-)_{I=1} \rangle | \pi^0 \rangle (p_{\pi^+} - p_{\pi^-}) \cdot p_{\pi^0} 
\rightarrow M_1(s) \frac{u-t}{2} \,, \nonumber \\
&& |(\pi^+\pi^0)_{I=1} \rangle  | \pi^- \rangle (p_{\pi^+} - p_{\pi^0}) \cdot p_{\pi^-} 
\rightarrow M_1(u) \frac{s-t}{2} \,, \nonumber \\
&& |(\pi^-\pi^0)_{I=1} \rangle  | \pi^+ \rangle (p_{\pi^0} - p_{\pi^-}) \cdot p_{\pi^+} 
\rightarrow M_1(t) \frac{u-s}{2} \,, 
\end{eqnarray}
\normalsize
which contribute to the amplitudes with total $I=0,1$, and $2$, and 
\begin{eqnarray}
&& |(\pi^+\pi^-)_{I=2} \rangle | \pi^0 \rangle 
\rightarrow M_2(s)  \,, \nonumber\\
&& |(\pi^+\pi^0)_{I=2} \rangle  | \pi^- \rangle 
\rightarrow M_2(u)  \,, \nonumber \\
&& |(\pi^-\pi^0)_{I=2} \rangle  | \pi^+ \rangle 
\rightarrow M_2(t)  \,, 
\end{eqnarray}
which contribute to the total $I=1$ and $2$ amplitudes. 
Using the Clebsch-Gordan coefficients tabulated in Ref.~\cite{Tanabashi:2018}, we find, after 
redefining $M_1/2\sqrt{2} \to M_1$ and $M_2 \sqrt{3/10}  \to M_2$, that 
$M_1^C (s,t,u)$, $M_0^{\slash{\!\!\!\!C}} (s,t,u)$, and $M_2^{\slash{\!\!\!\!C}} (s,t,u)$ are 
precisely as given in Eqs.~(\ref{anileut})~\cite{Anisovich:1996tx}, (\ref{CPV_isospin0}), and (\ref{CPV_isospin2}).  
Note that only the $C$-odd amplitudes are odd under $t \leftrightarrow u$ as needed. 
Adding the possible total $I$ amplitudes in leading order in $C$, $CP$, and $I$ violation, 
with their associated coefficients, yields Eq.~(\ref{total_amplitude}). 

We now turn to 
our isospin decomposition of the 
$\eta \to \pi^+ \pi^- \pi^0$ amplitude of Gasser and Leutwyler
through ${\cal O}(p^4)$~\cite{Gasser:1984pr}:
\begin{eqnarray}
  M_0 (s)
   &=&\left[ \frac{2 (s - s_0)}{\Delta} + \frac{5}{3} \right]
  \frac{1}{2 F_{\pi}^2} (2 s - M_{\pi}^2) J^r_{\pi \pi} (s) \nonumber\\
  &+& \frac{1}{6 F_{\pi}^2 \Delta} (4 M_K^2 - 3 M_{\eta}^2 - M_{\pi}^2) (s
  - 2 M_{\pi}^2) J_{\pi \pi}^r (s)\nonumber\\
  & + &  \frac{1}{4 F_{\pi}^2 \Delta} \Big[ - 6 s^2 + s (5 M_{\pi}^2 + 4
  M_K^2 + 3 M_{\eta}^2) \nonumber\\
  &&~~~~~~~~~ - 4 M_K^2 ( M_{\eta}^2 + \frac{1}{3} M_{\pi}^2
  ) \Big] J_{K K}^r(s)\nonumber\\
  & + &  \frac{M_{\pi}^2}{3 F_{\pi}^2 \Delta} \left( 2 s - \frac{11}{3}
  M_{\pi}^2 + M_{\eta}^2 \right) J_{\pi \eta}^r(s)  - \frac{M_{\pi}^2}{2 F_{\pi}^2} J_{\eta \eta}^r(s)\nonumber\\
  &-  &   \frac{3 s}{8 F_{\pi}^2} \frac{(3 s - 4 M_K^2)}{(s - 4 M_K^2)}
\Big(J^r_{KK}(s) - J^r_{KK}(0)-\frac{1}{8\pi^2}\Big)\nonumber\\
  & + &   \left[ 1 + a_1 + 3 a_2\Delta + a_3 (9 M_{\eta}^2 - M_{\pi}^2) + \frac{2}{3} d_1\right.\nonumber\\
  &&~~\left.   + \frac{8
  M_{\pi}^2}{3 \Delta} d_2 \right]\left( 1 + 3 \frac{s - s_0}{\Delta} \right) + a_4 - \frac{8}{3}\frac{M_{\pi}^2}{\Delta} d_1 \nonumber\\
  &-& \frac{3}{\Delta}( 2 {\mu}_{\pi} +  {\mu}_K )(s-s_0)\nonumber\\
  &+& \Big(\frac{4L_3}{F_0^2\Delta} - \frac{1}{64\pi^2 F_\pi^2 \Delta} \Big)
\Big(\frac{4}{3}s^2 - 9s_0s + 9s_0^2 \Big)\nonumber\\
    & -& \frac{1}{64\pi^2 F_\pi^2 \Delta}3(s-s_0)(4M_\pi^2 + 2M_K^2) \,,\label{M_0(s)}
\end{eqnarray}
\begin{eqnarray}
  M_1 (z)  & = &  \frac{1}{4 \Delta F_{\pi}^2} \Big[ (z - 4 M_{\pi}^2) 
J^r_{\pi \pi}   (z) \nonumber\\
&& + \Big( \frac{1}{2} z - 2 M_K^2 \Big) J_{K K}^r (z) \Big]\,,\label{M_1(s)}
\end{eqnarray}
and 
\begin{eqnarray}
  M_2 (z) 
  & = & \left( 1 - \frac{3}{2} \frac{z - s_0}{\Delta} \right) \left[ -
  \frac{1}{2 F_{\pi}^2} (z - 2 M_{\pi}^2) J^r_{\pi \pi} (z) \right.\nonumber\\
  &&\left.+ \frac{1}{4F_{\pi}^2} (3 z- 4 M_K^2) J_{K K}^r (z) 
+ \frac{M_{\pi}^2}{3 F_{\pi}^2} J_{\pi \eta}^r (z) \right] \nonumber\\
  &  &+ \left(\frac{1}{64\pi^2 F_\pi^2 \Delta} - \frac{4L_3}{F_0^2\Delta}\right)z^2 \,,\label{M_2(s)}
  \end{eqnarray}
where 
$\Delta=M_\eta^2 - M_\pi^2$, 
$M_\pi^2=(2M_{\pi^+}^2 + M_{\pi^0}^2)/3$, and $M_K^2 = (M_{K^+}^2 + M_{\bar{K}^0}^2)/2$. 
We refer to Ref.~\cite{Gasser:1984gg} 
for 
$J^r_{P Q} (z)$, noting Eqs.~(8.8-8.10) and (A.11), 
where P and Q denote the mesons 
$\pi$, $K$, or $\eta$, and to Ref.~\cite{Gasser:1984pr}
for $a_i$ and $d_i$. We note that the $J^r_{P Q} (z)$ carry renormalization-scale $\mu$
dependence, though cancelling that dependence is beyond the scope of
our current approach---we note a similar issue arises 
in the use of
the pion form factor in the analysis of $B\to \pi (\rho \to \pi\pi)$
decay~\cite{Gardner:2001gc}. 
For this choice of $M_{\pi}$ and the use of physical phase space we need to evaluate
the possible two-particle thresholds with care. The rescattering function 
$J^r_{\pi\pi}(z)$ contains $\sigma(s) = \sqrt{1-4m_\pi^2/z}$. 
If we use $m^2_\pi = M_{\pi}^2$, then 
for $M_I(z)$ with $z=t$ or $u$ evaluated at its minimum value the argument
of the square root is less than zero. 
To avoid this problem, we use $\sigma(z) = \sqrt{1-(M_{\pi^\pm} + M_{\pi^0})^2/z}$ 
for $z=t$ or $u$. 
For $M_I(s)$, though, $s_{\rm min}=4M_{\pi^+}^2 $ and this problem does not occur. 
However, for consistency we use $\sigma(s) = \sqrt{1-4M_{\pi^+}^2/s}$ 
for $M_I(s)$. 
Moreover, we note
$J^r_{\pi\eta}(s)$ contains $\nu(s)=\sqrt{(s-(M_\eta-m_\pi)^2)(s-(M_\eta+m_\pi)^2)}$. 
If we use $m_\pi = M_\pi$, then 
for $M_I(s)$ at the maximum of $s$, we once again find the argument of the square root
to be less than zero. 
To avoid this, we use  $\nu(s)=\sqrt{(s-(M_\eta-M_{\pi^0})^2)(s-(M_\eta+M_{\pi^0})^2)}$ 
for $M_I(s)$. To be consistent, we 
use $\nu(z)=\sqrt{(z-(M_\eta-M_{\pi^+})^2)(z-(M_\eta+M_{\pi^+})^2)}$ 
for $M_I(z)$ with $z=t$ or $u$. As a check of our assessments we have
extracted  the $C$- and $CP$-conserving Dalitz plot parameters from this amplitude. 
Describing the $CP$-conserving piece of $|A(s,t,u)|^2$ 
by $N(1+aY+bY^2+dX^2+fY^3+gX^2 Y)$, recalling Eq.~(\ref{amplitude_expansion}), 
we find using a global fit that 
$a=-1.326$, $b=0.426$, $d=0.086$, $f=0.017$, and $g=-0.072$. These results 
compare favorably to the global fit results of Ref.~\cite{Albaladejo:2017hhj}; 
namely, $a=-1.328$, $b=0.429$, $d=0.090$, $f=0.017$, and $g=-0.081$. That work also 
uses the decay amplitude of Ref.~\cite{Gasser:1984pr}
through ${\cal O}(p^4)$ and the same value 
of $L_3$ but includes electromagnetic corrections through ${\cal O}(e^2p^2)$ as well. 

In evaluating the BG amplitude~\cite{Bijnens:2007pr}
we note that an overall 2 should not 
appear on the second right-hand side of Eq.(3.23); this is needed
for the result to agree with that of Ref.~\cite{Gasser:1984pr}. 

Values of the strong functions associated with the $CP$-violating parameters
$\text{Re}(\alpha)$, $\text{Im}(\alpha)$, $\text{Re}(\beta)$, $\text{Im}(\beta)$
in Eq.~(\ref{CPV_interference})
on our analysis grids in $(X,Y)$ are available upon request.


\bibliography{patterns_0423.bbl}

\end{document}